\documentclass[iop]{emulateapj}

\usepackage{amssymb,amsmath,natbib,graphicx}
\begin{document}

\newcommand{\hi}          {\mbox{\rm H{\small I}}}
\newcommand{\hii}         {\mbox{\rm H{\small II}}}
\newcommand{\htwo}        {\mbox{H$_{2}$}}
\newcommand{\jone}        {\mbox{$J=1-0$}}
\newcommand{\jtwo}        {\mbox{$J=2-1$}}
\newcommand{\jthree}        {\mbox{$J=3-2$}}
\newcommand{\jfour}        {\mbox{$J=4-3$}}
\newcommand{\um}          {$\mu$m}
\newcommand{\ha}          {H$\alpha$}
\newcommand{\kmpers}      {\mbox{\rm km~s$^{-1}$}}
\newcommand{\percmcu}     {\mbox{\rm cm$^{-3}$}}
\newcommand{\msun}        {\mbox{\rm M$_\odot$}}
\newcommand{\msunperpcsq} {\mbox{\rm M$_\odot$~pc$^{-2}$}}
\newcommand{\msunperpcsqyr} {\mbox{\rm M$_\odot$~pc$^{-2}$~yr$^{-1}$}}
\newcommand{\msunperyr}   {\mbox{\rm M$_\odot$~yr$^{-1}$}}
\newcommand{\msunperpccu} {\mbox{\rm M$_\odot$~pc$^{-3}$}}
\newcommand{\msunperyrkpcsq} {\mbox{\rm M$_\odot$~yr$^{-1}$~kpc$^{-2}$}}
\newcommand{\xco}         {\mbox{$X_{\rm CO}$}}
\newcommand{\xcot}         {\mbox{$X_{\rm CO,20}$}}
\newcommand{\aco}         {\mbox{$\alpha_{\rm CO}$}}
\newcommand{\xcounits}    {\mbox{\rm cm$^{-2}$(K km s$^{-1}$)$^{-1}$}}
\newcommand{\acounits}  {\mbox{\rm M$_\odot$ (K km s$^{-1}$ pc$^2$)$^{-1}$}}
\newcommand{\Lcounits}  {\mbox{\rm K km s$^{-1}$ pc$^2$}}
\newcommand{\Kkmpers}     {\mbox{\rm K km s$^{-1}$}}
\newcommand{\Kkmperspcsq} {\mbox{\rm K km s$^{-1}$ pc$^2$}}
\newcommand{\co}          {\mbox{$^{12}$CO}}
\newcommand{\cothree}          {\mbox{$^{13}$CO}}
\newcommand{\Ico}         {\mbox{I$_{\rm CO}$}}
\newcommand{\av}          {\mbox{$A_V$}}
\newcommand{\percmsq}     {\mbox{cm$^{-2}$}}
\newcommand{\cii}         {\mbox{\rm [C{\small II}]}}
\newcommand{\nii}         {\mbox{\rm [N{\small II}]}}
\newcommand{\ci}         {\mbox{\rm [C{\small I}]}}
\newcommand{\wco}         {\mbox{\rm W(CO)}}
\newcommand{\fscii}       {($^2$P$_{3/2}\rightarrow^2$P$_{1/2}$)}
\newcommand{\Smol}	  {\mbox{$\Sigma_{\rm mol}$}}
\newcommand{\Mmol}	  {\mbox{${\rm M}_{\rm mol}$}}
\newcommand{\Ssfr}	  {\mbox{$\Sigma_{\rm SFR}$}}
\newcommand{\Sgmc}	  {\mbox{$\Sigma_{\rm GMC}$}}
\newcommand{\Lco}	  {\mbox{$L_{\rm CO}$}}
\newcommand{\ag}{\mbox{ \raisebox{-.4ex}{$\stackrel{\textstyle >}{\sim}$} }}
\newcommand{\al}{\mbox{ \raisebox{-.4ex}{$\stackrel{\textstyle <}{\sim}$} }}
\newcommand{\dgr}         {\mbox{$\delta_{\rm DGR}$}}
\newcommand{\dgrp}         {\mbox{$\delta_{\rm DGR}'$}}
\newcommand{\rco}{R_{\rm CO}}
\newcommand{\rht}{R_{\rm H_2}}
\newcommand{\davdg}{\Delta A_{V}}
\newcommand{\Ropt}{\mbox{$R_{\rm 1/2,Opt}$}}
\newcommand{\Rco}{\mbox{$R_{\rm 1/2,CO}$}}

\input epsf.tex

\input epsf.def

\input psfig.sty

\title{High Resolution Imaging of PHIBSS $z\sim2$ Main Sequence Galaxies in CO $J=1\rightarrow0$} 

\author{A. D. Bolatto\altaffilmark{1}, S. R. Warren\altaffilmark{1}, A. K. Leroy\altaffilmark{2}, L. J. Tacconi\altaffilmark{3}, N. Bouch\'e\altaffilmark{4,16}, N. M. F\"orster Schreiber\altaffilmark{3}, R. Genzel\altaffilmark{3,14,15}, M. C. Cooper\altaffilmark{5}, D. B. Fisher\altaffilmark{6}, F. Combes\altaffilmark{7}, S. Garc\'{\i}a-Burillo\altaffilmark{8}, A. Burkert\altaffilmark{3,9}, F. Bournaud\altaffilmark{10}, A. Weiss\altaffilmark{11}, A. Saintonge\altaffilmark{12}, S. Wuyts\altaffilmark{3}, \& A. Sternberg\altaffilmark{13}}

%F. Combes, S. Garc\'{\i}a-Burillo, R. Neri, D. Lutz, A. Saintonge, S. Berta, B. Magnelli, T. Contini, S. Lilly, J. Boissier, F. Boone, N. Bouch\'e, F. Bournaud, A. Burket, M. Carollo, L. Colina, P. Cox, C. Feruglio, J. Freundlich, J. Graci\'a-Carpio, S. Juneau, K. Kovac, M. Lippa, T. Naab, P. Salome, A. Renzini, A. Sternberg, F. Walter, B. Weiner, A. Weiss, \& S. Wuyts

\altaffiltext{1}{Department of Astronomy and Joint Space Institute, University of Maryland, College Park, MD 20642, USA}

\email{bolatto@astro.umd.edu}

\altaffiltext{2}{Department of Astronomy, Ohio State University, Columbus, OH 43210, USA}

\altaffiltext{3}{Max-Planck-Institut f\"ur Extraterrestrische Physik (MPE), Giessenbachstr., D-85748 Garching, Germany}

\altaffiltext{4}{CNRS/IRAP, 14 Avenue E. Belin, F-31400 Toulouse, France}

\altaffiltext{5}{Department of Physics \& Astronomy, Frederick Reines Hall, University of California, Irvine, CA 92697}

\altaffiltext{6}{Centre for Astrophysics and Supercomputing, Swinburne University of Technology, P.O. Box 218, Hawthorn, VIC 3122, Australia}

\altaffiltext{7}{Observatoire de Paris, LERMA, CNRS, 61 Av. de l'Observatoire, F-75014 Paris, France}

\altaffiltext{8}{Observatorio Astron\'omico Nacional-OAN, Observatorio de Madrid, Alfonso XII, 3, 28014 - Madrid, Spain}

\altaffiltext{9}{Universit\"atssternwarte der Ludwig-Maximiliansuniversit\"at , Scheinerstr. 1, D-81679 M\"unchen, Germany}

\altaffiltext{10}{Service d'Astrophysique, DAPNIA, CEA/Saclay, F-91191 Gif-sur-Yvette Cedex, France}

\altaffiltext{11}{Max Planck Institut f\"ur Radioastronomie (MPIfR), Auf dem H\"ugel 69, 53121 Bonn, Germany}

\altaffiltext{12}{Department of Physics \& Astronomy, University College London, Gower Place, London WC1E 6BT, UK}

\altaffiltext{13}{School of Physics and Astronomy, Tel Aviv University, Tel Aviv 69978, Israel}

\altaffiltext{14}{Department of Physics, Le Conte Hall, University of California, Berkeley, CA 94720, USA} 

\altaffiltext{15}{Department of Astronomy, Campbell Hall, University of California, Berkeley, CA 94720, USA}

\altaffiltext{16}{University Paul Sabatier of Toulouse/ UPS-OMP/ IRAP, F-31400 Toulouse, France}

\shorttitle{High Resolution CO \jone\ on $z\sim2$ Main Sequence Galaxies}
\shortauthors{A. D. Bolatto et al.}

\begin{abstract}
We present Karl G. Jansky Very Large Array observations of the CO \jone\
transition in a sample of four $z\sim2$ main sequence galaxies. These
galaxies are in the blue sequence of star-forming galaxies at their
redshift, and are part of the IRAM Plateau de Bure HIgh-$z$ Blue
Sequence Survey (PHIBSS) which imaged them in CO
\jthree. Two galaxies are imaged here at high signal-to-noise, allowing
determinations of their disk sizes, line profiles, molecular surface
densities, and excitation. Using these and published measurements, we
show that the CO and optical disks have similar sizes in main-sequence
galaxies, and in the galaxy where we can compare CO \jone\ and
\jthree\ sizes we find these are also very similar. Assuming a Galactic
CO-to-\htwo\ conversion, we measure surface densities of
$\Smol\sim1200$~\msunperpcsq\ in projection and estimate
$\Smol\sim500-900$~\msunperpcsq\ deprojected. Finally, our data yields
velocity-integrated Rayleigh-Jeans brightness temperature line ratios
$r_{31}$ that are approximately unity. In addition to the similar disk
sizes, the very similar line profiles in \jone\ and
\jthree\ indicate that both transitions sample the same kinematics,
implying that their emission is coextensive.  We conclude that in
these two main sequence galaxies there is no evidence for significant
excitation gradients or a large molecular reservoir that is diffuse or
cold and not involved in active star-formation.  We suggest that
$r_{31}$ in very actively star-forming galaxies is likely an indicator
of how well mixed the star formation activity and the molecular
reservoir are.
\end{abstract}


\section{Introduction}

Galaxy formation and evolution are regulated by the interplay
between the hierarchical merging of dark matter halos, the accretion
of primordial and recycled gas, the transport of gas within galaxy
disks, and the subsequent gravitational fragmentation, phase
transition, and formation of molecular clouds. The star formation that
is the end-point of this process gives rise to nucleosynthesis, and
metal-enriched outflows driven by stellar winds, radiation pressure,
supernovae, and AGN activity \citep[e.g.,][]{DAVE2011a,DAVE2011b}. These
mechanisms interact in complex ways, giving rise to the rich
phenomenology observed in galaxies today.

Surveys at $z\sim1-2$ have shown that most star-forming galaxies (SFGs)
populate a tight main sequence \citep{NOESKE2007,DADDI2007,RODIGHIERO2010},
suggesting that the evolution of a typical galaxy is regulated by
secular processes related to gas accretion from the cosmic web, with
subsequent star formation, galactic outflows, and reaccretion of gas
\citep[e.g.,][]{OPPENHEIMER2010,BOUCHE2010,DAVE2012,LILLY2013,FORBES2014}. The gas accretion is provided by a combination of
smooth and lumped accretion, the latter likely in the form of minor
mergers \citep[e.g.,][]{KERES2005,DEKEL2009b,BROOKS2009}.  In this
paradigm major mergers play a secondary role at forming the majority
of galaxies, but are still critical for creating the most massive and
the most actively star-forming galaxies. Until recently, almost all of
the observations of gas in galaxies at high redshift targeted the rare
``behemoths'' \citep[see][]{CARILLI2013}, typically quasar hosts or merging
galaxies bright at submillimeter wavelengths \citep[or even merging
groups of galaxies,][]{IVISON2013}. The exceptions were a few
intrinsically faint but highly lensed objects, where lensing
complicates the data interpretation
\citep{BAKER2004,COPPIN2007,RIECHERS2010}. The increased sensitivity
in the millimeter-wave regime makes it now possible to directly
observe normal massive star-forming galaxies at high redshift. Indeed,
high-$z$ main sequence SFGs are considerably
more active in star formation than similar mass galaxies at $z=0$,
likely because they are richer in gas than present day blue sequence
galaxies, as found by molecular gas observations
\citep{TACCONI2010,TACCONI2013,DADDI2010a,DADDI2010b}.

Early universe molecular gas surveys usually target the mid-$J$
transitions of the CO molecule, which are accessible for $1<z<3$ in
the transparent $1-3$~mm atmospheric windows. Additionally, for a
fixed gas mass and typical conditions these transitions are more
luminous and easier to detect than the ground level transition. This,
however, introduces an additional source of uncertainty when
converting their fluxes to \htwo\ masses at high-$z$
\citep{WEISS2007,CARILLI2013}. Excitation data remain scarce for
high-$z$ SFGs. The lowest galaxy luminosities currently probed by high
redshift CO observations correspond to the high luminosity end of the
main sequence, and the data suggest their excitation is intermediate
between that observed in high-z Submillimeter Galaxies (SMGs) or
local Ultra-Luminous IR Galaxies (ULIRGs), and the Milky Way
\citep{FIXSEN1999}. Observations find 
CO main-beam velocity-integrated Rayleigh-Jeans brightness temperature
line ratios $r_{31}\equiv T_{mb}\Delta v (J=3-2)/T_{mb}\Delta v
(J=1-0) \sim0.3-0.8$ in a handful of high-$z$ main sequence SFGs
\citep{DANNERBAUER2009,ARAVENA2010,RIECHERS2010,ARAVENA2014}. 
The excitation picture, however, is not clear cut. Studies of samples
of local and high-$z$ luminous and ultraluminous IR galaxies also
suggests a typical $r_{31}\sim0.6$ with very large dispersions
\citep{HARRIS2010,PAPADOPOULOS2011,IVISON2011,GREVE2014}.

The observed $r_{31}$ in high-$z$ main sequence SFGs can be reproduced
by a range of conditions. The typical conditions found in the highly
idealized single-component models are $T_{kin}\approx90$~K with
$n\approx600$~\percmcu\ to $T_{kin}\approx10$~K with
$n\approx2500$~\percmcu, together with gas filling factors from $2\%$
to $8\%$ \citep{ARAVENA2010}.  This interpretation is by no means
unique, however, as realistic galaxies contain multiple
components, and different transitions may not be necessarily
coextensive --- for example if dense warm star-forming regions were
surrounded by cooler, less dense envelopes of gas
\citep[e.g.][]{HARRIS2010}. It is also possible to hide a significant
amount of mass in a cold component that would be hard to detect
\citep{PAPADOPOULOS2012}. The inclusion of $J>3$ transitions reveals
that, at least in some of the main sequence SFGs, the ``one
component'' picture of the ISM is too simplistic, although the data
support the idea of a somewhat lower excitation in main sequence
galaxies than in SMGs \citep{DADDI2014,BOURNAUD2014a}. In particular,
the $T_{kin}$ and $n$ inferred above would produce too little, if any,
$J>4$ emission in contrast with the observations. Better sensitivity
observations of the \jone\ transition, observations of transitions
with $J>3$, and higher spatial resolutions are necessary to
definitively constrain the molecular excitation in these galaxies.

There are a handful of resolved molecular measurements of $z\sim1-2$
main sequence galaxies. \citet{TACCONI2010} discusses in particular
\jthree\ observations of one $z\sim1.5$ SFG at a resolution of
$\theta\sim0\farcs65$ that reveal a collection of clumps with gas
masses of $\sim5\times10^9$~\msun, sizes of $\sim2-4$~kpc, and surface
densities $\ge500$~\msunperpcsq. \citet{TACCONI2013} expands on these
results, determining $\sim20$ CO \jthree\ sizes (mostly for $z\sim1$
objects), reaching the conclusion that the UV/optical disk size and
the CO size are similar. \citet{GENZEL2013} examine high-resolution
observations of the $z\sim1.5$ galaxy EGS13011166, a disky system with
massive clumps and a globally unstable gas disk.
\citet{FREUNDLICH2013} take advantage of
position-velocity information to apportion the \jthree\ emission from
four $z\sim1.5$ SFGs in clumps, inferring typical gas surface densities
$\Smol\sim400$~\msunperpcsq\ with a large dispersion.
\citet{ARAVENA2014} presents a few \jone\ observations
of $z\sim1-2$ SFGs, including an earlier dataset for the $z=2.2$
galaxy BX610, one of the sources studied in this paper. In the
$z\sim1.5$ SFG that they observe at high resolution they find clumps
with masses $\sim4-9\times10^9$~\msun\ that make up 40\% of the
emission. Giant clumps are a natural consequence of the gas richness
of these high-$z$ systems, which makes their disks unstable
\citep[e.g.,][]{DEKEL2009a,GENZEL2013,BOURNAUD2014b}. A clumpy morphology has
also been observed in the relatively rare, very gas-rich and strongly
star-forming local disks \citep{GREEN2014,FISHER2014}.

In this work we present new, very sensitive observations of the CO
\jone\ transition in main sequence SFGs at redshift 
$z\sim2.3$, during the peak of cosmic star formation. We discuss the
observations and data reduction in \S2. In \S3 we present and discuss
the results; particularly focusing on sizes, surface densities, and
excitation. In \S4 we summarize them and present the conclusions.

\section{Observations and Reduction}

We selected our sources as the brightest CO \jthree\ emitters at
$z\sim2.2-2.3$ in the IRAM Plateau de Bure HIgh-$z$ Blue Sequence
Survey (PHIBSS) sample \citep{TACCONI2010,TACCONI2013}. Original
optical data for these objects are discussed by
\citet{ERB2006}.  We observed the CO (J=$1-0$) transition at 115.2712
GHz rest frequency in Q1700MD94, Q2343BX610, Q1700MD69, and Q1700MD174
using the Karl G. Jansky Very Large Array (Jansky VLA) in the C- and
D-array configurations, with typical angular resolution of
$\sim0.7\arcsec$ and $\sim2\arcsec$ respectively. In order to observe
the redshifted CO line the observations employ the Ka-band receiver,
which covers the frequency range $26.5 - 40.0$ GHz.

C-array observations were carried out between 12 October 2010 and 09
January 2011 under project 10B-106.  These data were obtained during
the VLA to Jansky VLA upgrade transition period, resulting in a
reduced bandwidth. Each C-array observation has 0.5 MHz spectral
resolution and covers a 128 MHz bandwidth centered on the redshifted
frequencies of our galaxies (34.55 GHz, 35.90 GHz, 35.06 GHz, and
34.49 GHz for MD94, BX610, MD69, and MD174 respectively).  

The follow up D-array data were taken between 07 February 2013 and 07
April 2013 under project 13A-115.  These data have 1 MHz spectral
resolution with a total of 4 GHz bandwidth coverage. The time invested
is approximately 8 hours per source in C-array. The D-array follow up
spent $4.5$ hours on MD94 and $8.5$ hours on BX610. Because only
$\sim40\%$ of the visibilities in C-array have $uv$ distances
$\leq10^5$ k$\lambda$ (probing size scales $\gtrsim2\arcsec$), and the
sources were found to be extended in the original observations, the
sensitivity added by the D-array observations is very significant.

All data were processed using the Jansky VLA reduction
pipeline\footnote{The Jansky VLA reduction pipeline is available for
download at this url:
https://science.nrao.edu/facilities/vla/data-processing/pipeline}
version 1.2.0 using the Common Astronomy Software Applications package
(CASA) version 4.1. The fluxes employed for the flux calibrators were
applied in the pipeline following the prescribed fits by
\citet{PERLEY2013}.  We visually inspected each observation after the
calibration scripts were completed to flag a few remaining bad
baselines.  The source was then split off from each individual
observation and combined into a single data set for imaging. After
imaging with 50 km~s$^{-1}$ channel widths, we cleaned down to
$\sim1\sigma$ on a 5$\arcsec\times$5$\arcsec$ box centered on the
source. The image cubes span a range of $\pm1000$ km s$^{-1}$ around
the optical redshift of the source for MD94 ($z=2.336$) and BX610
($z=2.211$), while MD69 ($z=2.288$) and MD174 ($z=2.342$) were imaged
over a $\pm500$~\kmpers\ span because of the reduced bandwidth
available in the transition period during which the C-array data were
acquired. For the calculation of physical parameters we use a standard
cosmology \citep[H$_0=69.6$~km~s$^{-1}$~Mpc$^{-1}$, $\Omega_M=0.315$,
$\Omega_\Lambda=0.685$,][]{KOMATSU2014}.

\noindent{\it{\underline{BX610:}}} 
One C-array track was not included in the imaging because of strong
pattern noise (maybe due to interference or correlator problems), but
this problem persists at a lower level in the remainder of the C-array
data for this object.  The flux and gain calibrators for BX610 were
3C48 and J2346+0930, respectively.  At our observing frequency 3C48
has a flux density of $S_\nu=802$ mJy which was applied to our data to
fix the flux scale.  The resulting flux of the gain calibrator was
$470$ mJy in the C-array data and $325$ mJy in the D-array data. We
inserted a second calibrator in the D-array observations (J2330+110)
to check the quality of the calibration including phase transfer. The
recovered flux of the check calibrator is $660$~mJy, in excellent
agreement with the power-law interpolation between the 2~cm and 0.7~cm
fluxes listed in the VLA calibrator catalog for this object of
640~mJy. The synthesized beam sizes range from
$0\farcs83\times0\farcs63$ (C only, natural weighting) to
$1\farcs44\times1\farcs21$ (C+D, Briggs weighting with robust
parameter set to 0.5, henceforth ``robust weighting'') and
$2\farcs02\times1\farcs87$ (C+D, natural weighting). The 1$\sigma$
noise per 50 \kmpers\ channel for the robust C+D cube is 52 $\mu$Jy
beam$^{-1}$. The high resolution
\jthree\ data were obtained at PdBI as part of PHIBSS, and have a 
resolution $1\farcs5\times0\farcs7$.  

\noindent{\it{\underline{MD94:}}} Along with MD94, flux (3C286) and gain
(J1645+6330) calibrators were also observed.  3C286 has a flux density
of $\approx$1800 mJy at 34.55 GHz which was used to fix the flux scale
of the data. The derived flux density of our gain calibrator was $370$
mJy in the C-array data and $500$ mJy in the D-array data. We also
included a secondary calibrator in the observations
(J1716+6836). Applying the gain solution to this object recovers
$S_\nu\approx430$~mJy while the VLA calibrator catalog would lead us
to expect $\sim670$~mJy. However, given its listed fluxes this
calibrator appears to be a highly variable flat spectrum source, and
it is very likely that changes are due to intrinsic variability. The
synthesized beam sizes range from $0\farcs77\times0\farcs72$ (C only,
natural weighting) to $0\farcs95\times0\farcs89$ (C+D, robust
weighting) and $1\farcs70\times1\farcs55$ (C+D, natural
weighting). The 1$\sigma$ noise per channel for the robust C+D cube is
58 $\mu$Jy beam$^{-1}$.

\noindent{\it{\underline{MD174 \& MD69:}}} These sources were 
not clearly detected in C-array observations and were not followed
with D-array.  Both galaxies were flux calibrated with observations of
3C286 and antenna gains were corrected with observations of
J1645+6330. The resulting rms noise per 50 km s$^{-1}$ channel is
approximately 78 and 68 $\mu$Jy beam$^{-1}$ for MD174 and MD69,
respectively. The integrated intensity maps of both objects show
positive signal at the position of the source that, while formally
close to $3\sigma$ for MD69, is hard to distinguish from many nearby noise
peaks. The smoothed spectrum at the position of MD69 shows what may be
a marginally significant line. The \jone\ luminosities of these
objects in Table
\ref{tab:photometry} can be compared to their respective \jthree\
luminosities of $1.9$ and $1.1\times10^{10}$
\Kkmperspcsq\ \citep[][]{TACCONI2013}. The resulting ratios are
consistent with our conclusions for MD94 and BX610 in
\S\ref{sec:excitation}, but do not add significant information.

\section{Results and Discussion}

\begin{figure*}
\begin{center}
\includegraphics[width=0.85\linewidth]{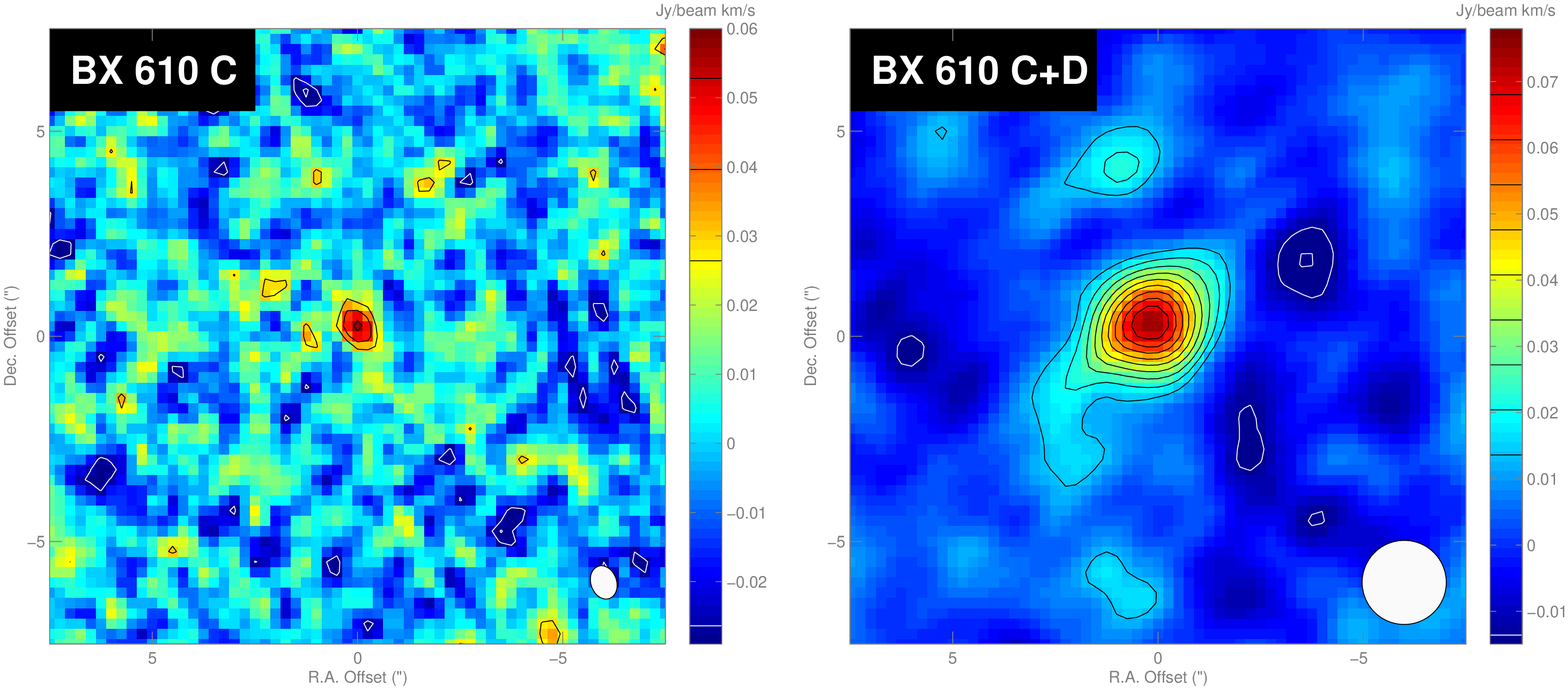}
\includegraphics[width=0.85\linewidth]{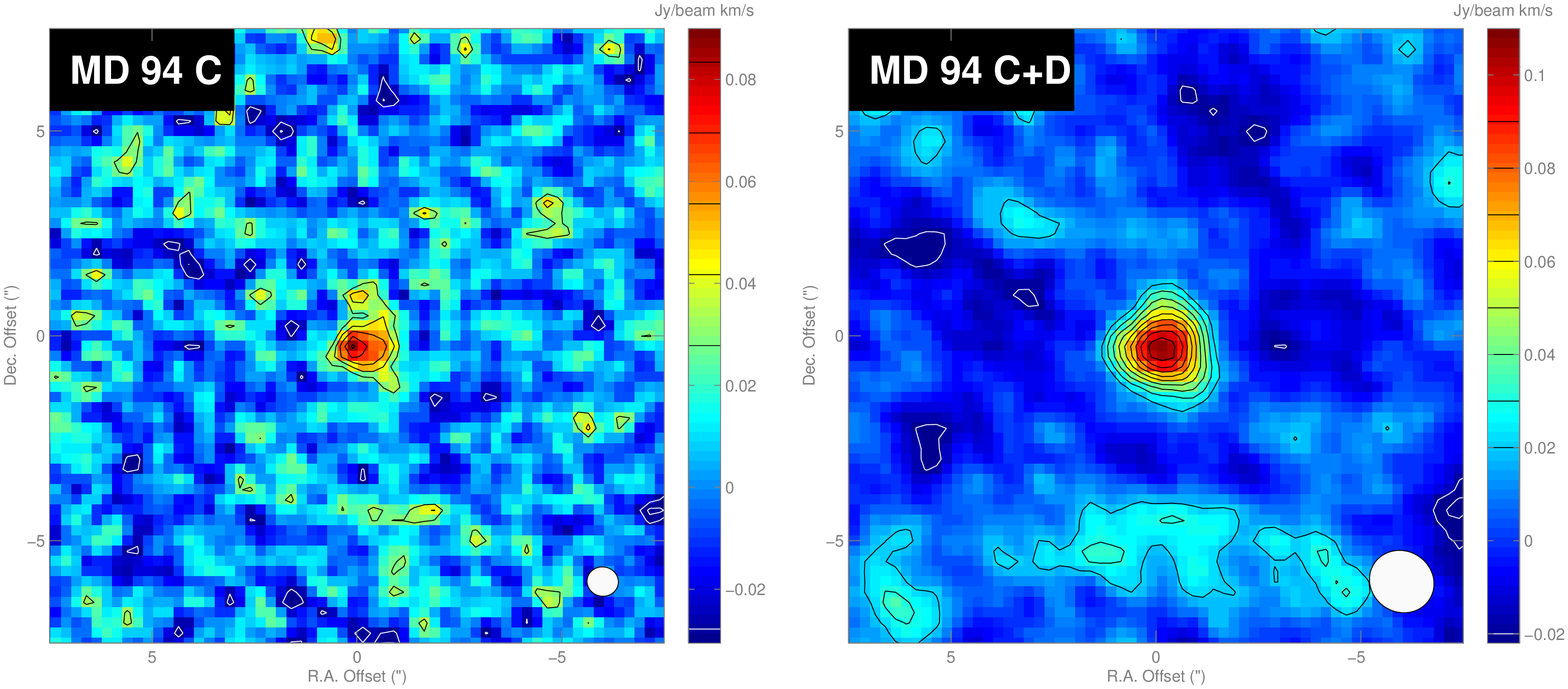}
\includegraphics[width=0.85\linewidth]{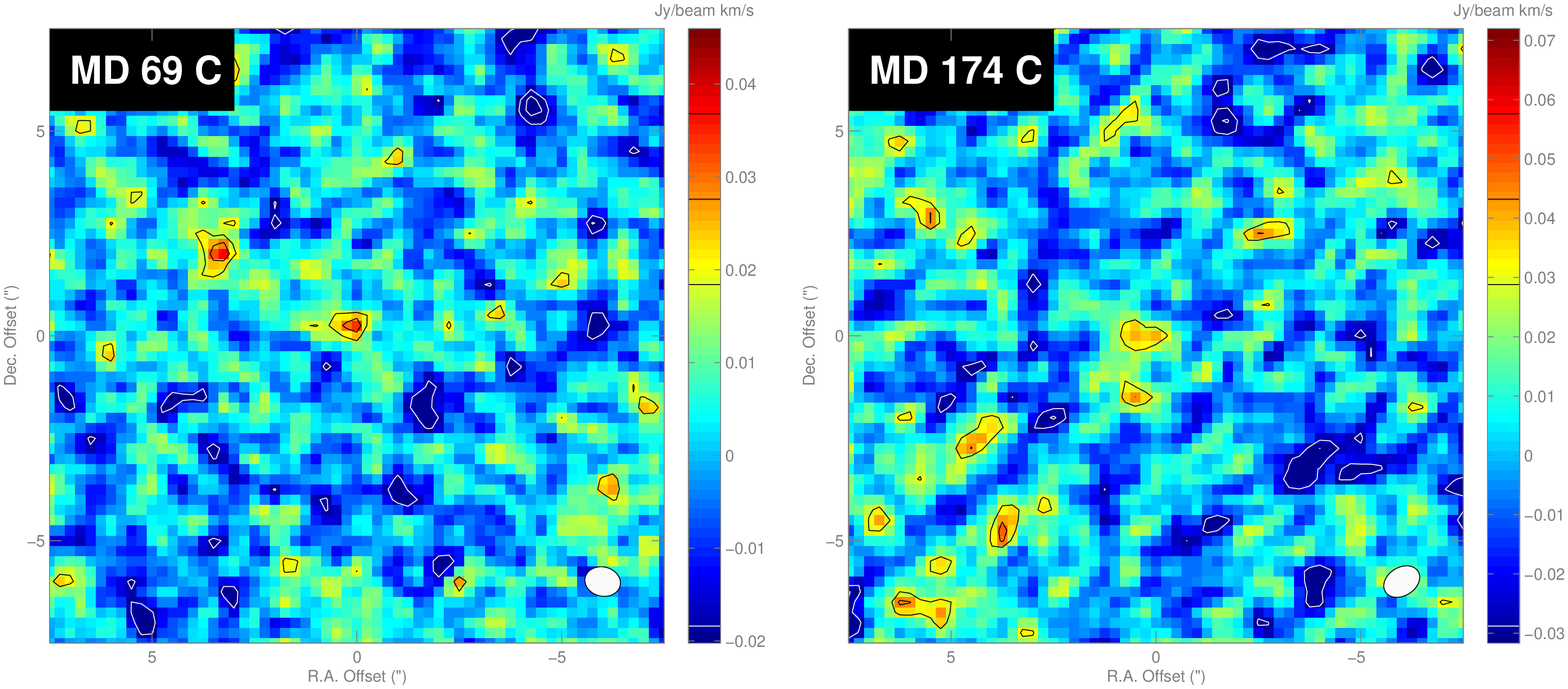}
\end{center}
\caption{CO \jone\ integrated intensity maps of our $z\sim2$ main-sequence targets. 
We show the naturally-weighted C-array observations for our four
galaxies, and the naturally weighted C+D combination for MD94 and
BX610.  The corresponding synthesized beam is illustrated in the
bottom-right corner. The contours start at $\pm2\sigma$ and increase
by $1\sigma$, with $\sigma$ listed in Table \ref{tab:photometry}
(white contours for negative values). The color bar to the right of
each panel indicates the intensity scale and the contour values used.\label{fig:maps}}
\end{figure*}

Both MD94 and BX610 are detected at good signal-to-noise, even in the
C-only data (Figure \ref{fig:maps}). MD69 and MD174, which are weaker
sources in the CO $3-2$ observations, were not clearly detected and
not followed up with D array observations, therefore we focus our
discussion on the previous two galaxies. Together with the earlier
publication of another BX610 dataset \citep{ARAVENA2014}, these
constitute the first CO \jone\ detections of unlensed $z>1.5$ main
sequence galaxies, and the first observations where measurements of
the intrinsic properties of the cold CO disks of main sequence
galaxies free of lensing corrections are possible at these redshifts.

Employing the parametrization of the main sequence of star-forming
galaxies derived for the AEGIS and COSMOS fields by
\citet{WHITAKER2012}, both MD94 and BX610 are very representative of
the massive end of the main sequence at their $z\sim2.3$ redshift.
Their specific star formation rates are sSFR$\sim1.8\times10^{-9}$ and
$2.1\times10^{-9}$ yr$^{-1}$ respectively, which at their inferred
stellar masses (M$_*\approx1.5\times10^{11}$ and
$1.0\times10^{11}$~\msun\ respectively) places them well within the main
sequence 0.3~dex scatter. Both are very active star-forming galaxies,
with respective SFR$\sim271\pm95$ and $212\pm74$~\msunperyr\
\citep{TACCONI2013}. MD94, the most massive of the pair, displays a
broad H$\alpha$ line and is classified as an AGN \citep{ERB2006}.
BX610 may contain an AGN, but not a very luminous one
\citep{FORSTERSCHREIBER2014,NEWMAN2014,GENZEL2014}. 
It is important to keep in mind that these galaxies were selected
because they are bright CO \jthree\ emitters, which makes it possible
to target them in CO \jone, and thus their CO properties may not be
representative of the population as a whole. Our CO measurements are
summarized in Table \ref{tab:photometry}, and the resulting physical
parameters in Table \ref{tab:parameters} together with relevant
parameters from \citet{TACCONI2013}.

\subsection{The Size of the Molecular Disks in High-$z$ Main Sequence Galaxies}

MD94 is clearly resolved in the observations (c.f., Table
\ref{tab:photometry}). The simple Gaussian fit (CASA {\tt imfit}) yields an elongated source
in the N-S direction, with deconvolved FWHM diameter (after averaging
the C-only and C+D robust results) of $\sim1\farcs3\times1\farcs0$ at
high significance ($\ge5\sigma$). With an image scale of $\sim8.4$ kpc
per arcsecond at this redshift, this corresponds to $\sim11\times8$
kpc. The residual RMS in the central region after removing the
Gaussian fit is very similar to that measured in empty regions of the
original image, showing that a 2D Gaussian is a good model for the
source at this signal-to-noise. BX610 is clearly resolved along its
major axis, with a deconvolved CO FWHM diameter of
$1\farcs6\times0\farcs6$ corresponding to $\sim13\times5$ kpc (for this source
we adopt the results of the C+D robust combination, which despite the
larger beam is better behaved than the C-only observation). We also
did UV-plane model fitting using the {\tt uvmodelfit} task in CASA,
and the results are essentially identical to those from image-plane
fitting.

In $z=0$ galaxies there is a good relation between the sizes of
molecular and stellar disks \citep[e.g.,][]{REGAN2001}.  The CO
emission is found in an approximately exponential disk with a $R_{\rm
CO}$ scale-length that is $R_{\rm CO}\simeq0.2\,R_{25}$, the isophotal
radius of the B-band light at 25$^{th}$ magnitude
\citep{YOUNG1991,YOUNG1995,SCHRUBA2011}. Consequently half of the CO
emission in local disks is contained within a radius
$\Rco=0.34\,R_{25}$. For optical light it is usually assumed that
the stellar disk scale length is $R_*\sim0.25\,R_{25}$ (implying an effective
radius containing half the optical light of $\Ropt\sim0.4R_{25}$), so
the stellar and molecular disks track each other very well in their
exponential part.

\begin{figure*}
\begin{center}
\includegraphics[width=3.5in]{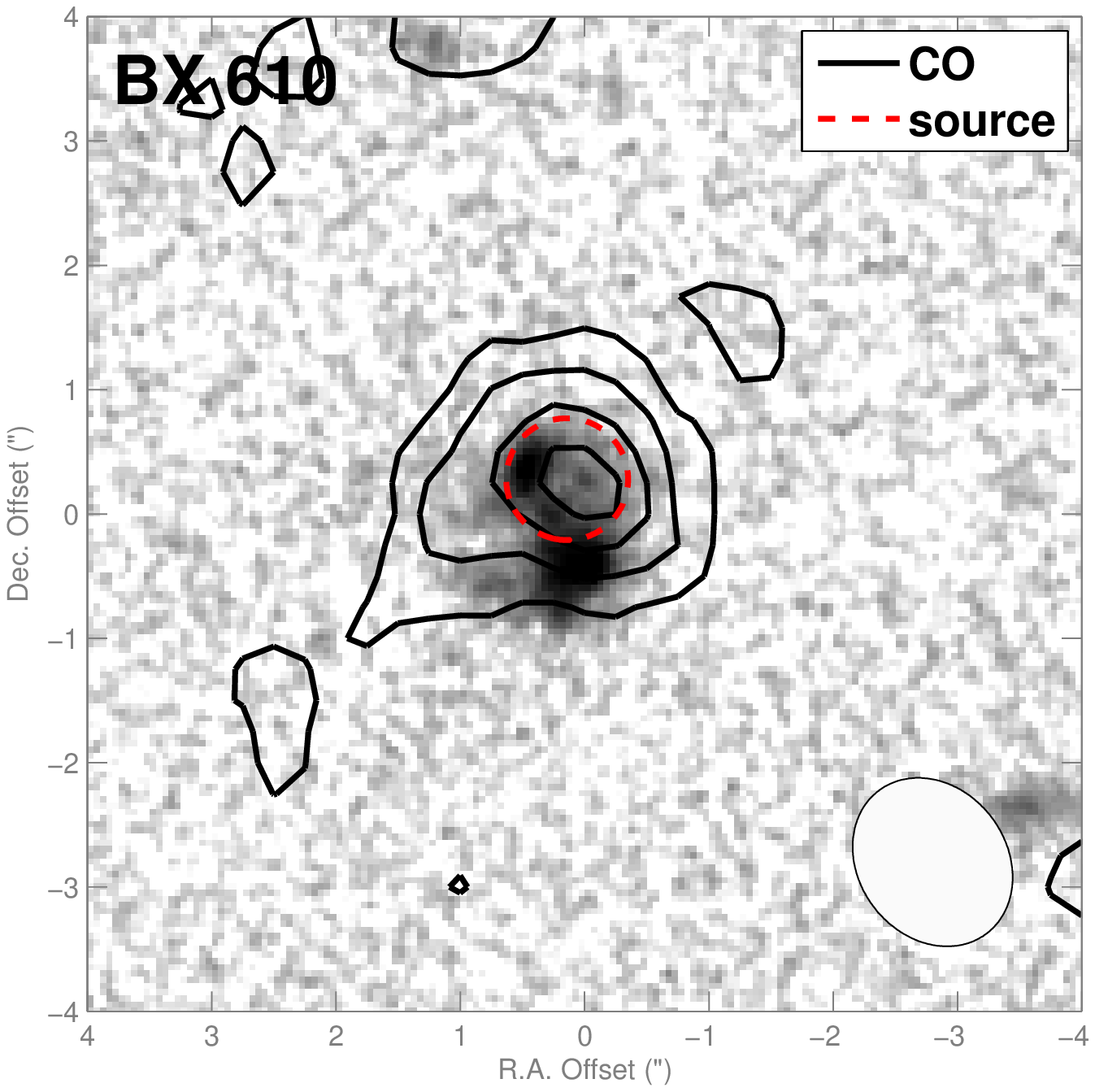}\includegraphics[width=3.5in]{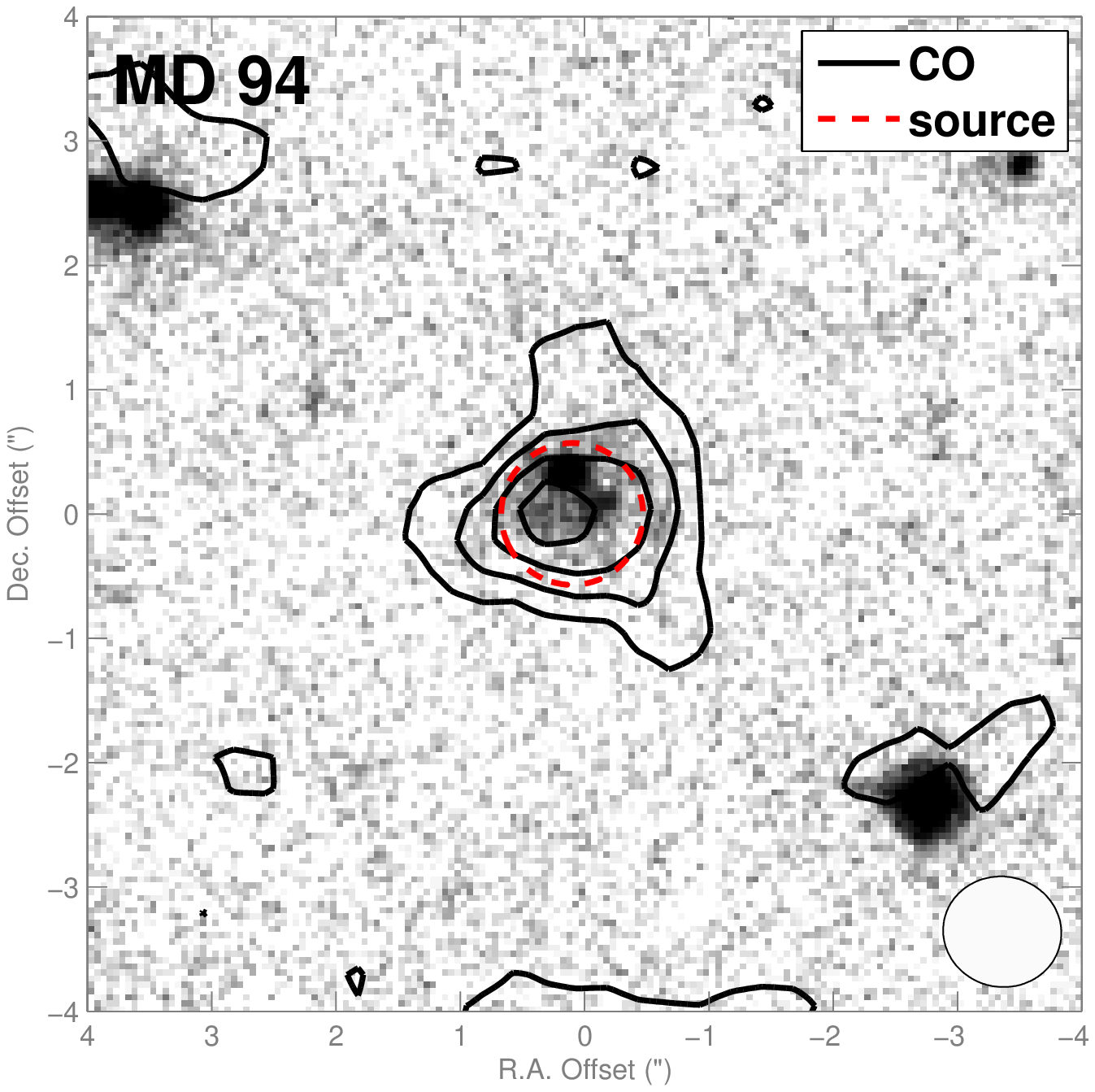}
\caption{
Comparison of CO and HST rest-frame optical or UV data. We show the
robust Briggs-weighted combination of C+D array CO \jone\ data in
contours (starting at $2\sigma$, in steps of $2\sigma$), overlaid on
the HST data. For BX610 the image is HST NICMOS 1.6 $\mu$m
observations from \citet{FORSTERSCHREIBER2011}, while for MD94 the
image is HST ACS 814 nm from \citet[][kindly provided by
A. Shapley]{PETER2007}. The stretch in these images is logarithmic to
emphasize the extended emission of the source disks. The corresponding
synthesized clean beam is illustrated in the bottom right corner. The
dashed circle illustrates the $\Rco$ for CO \jone, our best
measurement of the intrinsic molecular source size.
\label{fig:optical}}. 
\end{center}
\end{figure*}

How do the CO and optical sizes relate in our high-$z$ SFGs? For a
Gaussian source 50\% of the flux is contained within the FWHM,
therefore the measured $R_{\rm 1/2}$ CO radius for MD94 and BX610 is
respectively $\Rco\sim4.8$ and $4.1$ kpc. We can compare directly
these numbers with near-IR NICMOS observations of BX610 (Figure
\ref{fig:optical}) that sample the optical disk of that galaxy
\citep{FORSTERSCHREIBER2011}. Rest-frame 5000 \AA\ observations of
BX610 yield an effective radius of $\Ropt\approx4.6$~kpc, which within
the errors is very comparable to the \jone\ $\Rco$ we obtain. To
put disk optical sizes in the $z\sim0$ context, the typical half-light
optical radius of a $M_*\sim10^{11}$~\msun\ in the SDSS sample
analyzed by \citet{SHEN2003} is $\Ropt\sim4$~kpc.  The existing HST
observations for MD94 (Figure \ref{fig:optical}) are rest-frame 2500
\AA\ \citep{PETER2007}, which combined with the AGN presence makes it
difficult to obtain a reliable disk size
\citep[the 4.8~kpc optical radius reported by][is from H$\alpha$]{TACCONI2013}.
Surface photometry shows that 80\% of the 2500 \AA\ light is in an
exponential component with $\Ropt=0.37\pm0.08$ arcseconds
($3.1\pm0.7$~kpc) using the \citet{FISHER2008} methodology. Converting
these measurements to an optical size is, however, not straightforward
even in local galaxies \citep[which display a range of
scalings,][]{TAYLOR2005,MUNOZ-MATEOS2007}, much less in the poorly
characterized high-$z$ population. 

It is interesting to compare our results to some other resolved
measurements of high-$z$ SMGs. \citet{BOTHWELL2010} analyze resolved
observations of intermediate excitation CO \jthree\ and \jfour\ in
several such galaxies at $z\sim1.2$ to $z\sim2$. Their typical CO 50\%
emission radii (computed as the harmonic mean of the major and minor
axis) are $\Rco\sim 3$~kpc, although the observations span a range
of $\Rco\sim1-5$~kpc. Similar measurements are reported by
\citet{TACCONI2008} and \citet{ENGEL2010}, who find typical CO sizes
$\Rco\lesssim4$~kpc. \citet{SHARON2015} find lensing-corrected
source sizes of $\Rco\sim1.7$~kpc in \jthree\ and
$\Rco\sim2.0$~kpc in \jone\ for the two components of the
$z\sim2.7$ SMG they study. The $\Rco\sim4.1$ and 4.8 kpc sizes
measured here suggest that the CO \jone\ disks of main sequence
galaxies tend to be somewhat more extended than the region producing
mid-J CO emission in SMGs at comparable redshift. On the other hand,
resolved CO \jone\ measurements of five $z\sim2$ SMGs by
\citet{IVISON2011} find complex morphologies with radii of
$\gtrsim6$~kpc for four of the objects (the fifth is unresolved) and a
median FWHM linewidth of 540~\kmpers. These SMG CO \jone\ sizes and
linewidths are larger than those measured for the same objects in the
mid-J transitions, and also larger than the disk sizes we measure here.

\begin{figure*}
\begin{center}
\includegraphics[width=3in]{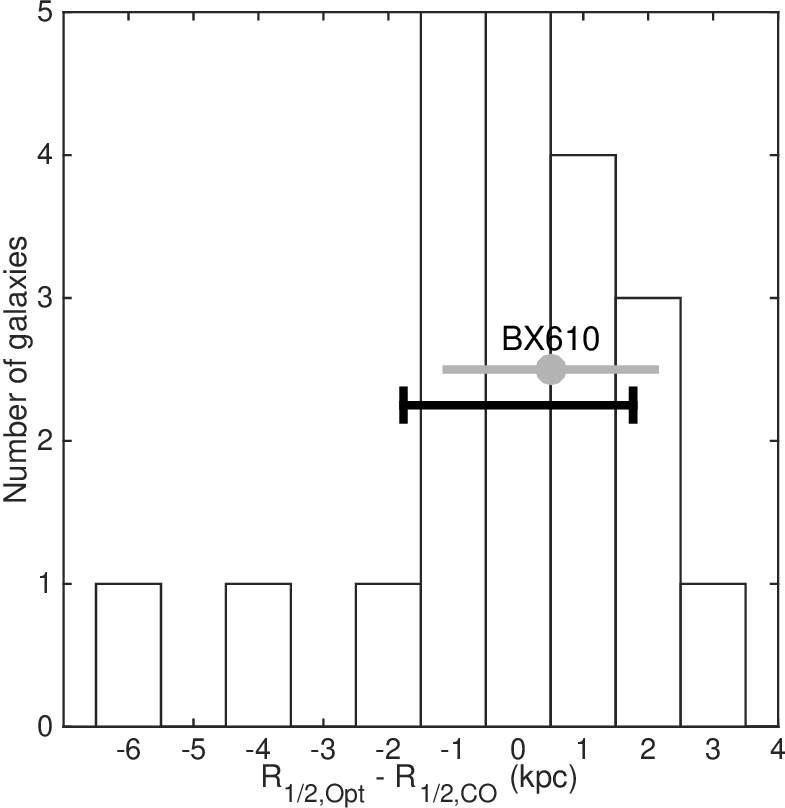}
\caption{
Comparison of optical and CO \jthree\ sizes in main-sequence galaxies
(mostly $z\sim1$) reported by \citet{TACCONI2013}. The histogram shows
the distribution of the difference between $R_{\rm 1/2}$ for the
optical light, \Ropt, and $R_{\rm 1/2}$ for the resolved CO
measurements (mostly from \jthree), \Rco. The black horizontal bar
illustrates the approximate $\pm1\sigma$ error in the difference (the
typical error in each size determination is $\sim1.3$~kpc). The gray
bar shows the \jone\ measurement presented here for BX610. The
distribution of the difference of optical and molecular sizes shows
that in the vast majority of these galaxies the CO and optical light
follow each other, with the exception of two galaxies that appear much
larger in CO (perhaps undergoing a
minor merger). \label{fig:tacconi_sizes}}
\end{center}
\end{figure*}

The similarity between the molecular and optical sizes in main
sequence galaxies is also seen in the CO \jthree\ measurements
reported by \citet{TACCONI2013}, mostly obtained for $z\sim1$ sources
(Fig. \ref{fig:tacconi_sizes}). These measurements suggest that
$z\sim2.3$ main sequence galaxy disks follow a scaling between optical
light and CO \jone\ similar to that in local disks. The CO is not
measurably more extended than the stars, as may occur if a significant
amount of pre-enriched newly-arrived material is present in the disk
outskirts. Such pre-enriched material is unlikely to arrive through
smooth accretion; it seems much more likely that occurs through
mergers, and in particular minor mergers given the placement of these
galaxies on the main sequence.  The fact that the scaling between
CO-emitting molecular gas and the stars is similar to that observed in
local disks suggests that we are not catching these $z\sim2$
main-sequence disks at an out-of-equilibrium stage. Rather, we observe
them in an equilibrium situation where the established stellar
population and the enriched CO-bright molecular gas that fuels the
current star formation have similar radial distributions.

\subsection{Molecular Masses and Surface Densities}

The excellent signal-to-noise of the observations for MD94 and BX610
permits, in principle, a precise determination of their CO \jone\
luminosity.  Unfortunately there is a distressing level of
inconsistency between the fluxes measured in the C and D
configurations, particularly for MD94, with the former being higher
than the latter. After detailed inspection of the data and calibrations
it is not clear what to attribute these differences to. In particular,
the changes in the flux measured for the sources between D and C
configurations are not mirrored by changes in the flux of their
respective gain calibrators which could suggest systematic problems
with the flux calibration (for MD94 the change is in the opposite
direction), and are not due to the phase noise as measured on the gain
calibrator (which shows RMS$\sim10-20$ degrees in either
configuration). After application of the calibrations we also recover
the expected flux for the secondary calibrator used in D-array, as
noted in \S 2. The flux inconsistency in MD94 is at the $\sim2\sigma$
level, which suggests that it could be simply due to noise, much
higher for the C-array data. The fact that the C-array observations
were obtained during a transition period for the instrument may also
have had an effect on the accuracy of those fluxes. Because of these
considerations we base our discussion on the fluxes from the naturally
weighted combination of C+D observations, which gives preferential
weighting to the D-array data and has the best signal-to-noise.

To compute the molecular masses we use Eq. 3 in \citet{BOLATTO2013},
which is applicable to CO \jone, with a Galactic CO-to-\htwo\
conversion factor. This conversion relies fundamentaly on two
assumptions: 1) the molecular gas is primarily self-gravitating, and
2) the ratio $\sqrt{\rho}/T$ between gas density, $\rho$, and
temperature, $T$, is roughly similar to that in local galaxy disks
\citep[][and references therein]{BOLATTO2013}. 
We use $\aco=4.36$~\acounits\ which includes the Helium contribution
to the mass as is customary ($\aco=3.2$~\acounits\ without the Helium
contribution). A number of arguments point to an approximately
Galactic value of the conversion factor for main sequence high-$z$
galaxies \citep{TACCONI2010,DADDI2010a,GENZEL2012}, including the
similarity between the gas masses inferred from CO and from dust
\citep{GENZEL2015}. This is supported by detailed numerical modeling of high-$z$
galaxies, which suggests that the average \aco\ in these systems is
not significantly different from \aco\ in local disks
\citep{BOURNAUD2014a}. We will see in \S\ref{sec:excitation} that
there are indications that the molecular gas temperature is higher in
these sources than the average in GMCs in the Galaxy. It is important
to remark that this does not automatically translate into a
lower-than-Galactic \aco. In a self-gravitating cloud what matters is
the $\sqrt{\rho}/T$ ratio, not just the temperature, as stated in
assumption number two above. Indeed, there are indications that
densities (certainly column densities) are much higher in these
objects than in Galactic GMCs.

Under these assumptions, the molecular masses inferred from CO \jone\
for MD94 and BX610 are $M_{\rm mol}\approx1.7\times10^{11}$ and
$1.1\times10^{11}$ M$_\odot$ respectively. These masses are similar to
the stellar masses determined for these galaxies \citep{TACCONI2013},
resulting in molecular gas fractions of
$f_{gas}\sim50\%$. \citet{BOLATTO2013} discuss a surface density
correction that is applicable if the CO-emitting gas is not
self-gravitating but bound to the overall potential of the galaxy as
is applicable in a merger and some galaxy centers (c.f., their
Eq. 31). For our galaxies, application of this correction would amount
to reducing their molecular mass by a factor of $\sim3$. This is
consistent with what detailed modeling recovers for very turbulent
starbursting mergers, but it is likely too large a correction for the
galaxies presented here \citep[by contrast modeling of very
actively star-forming clumpy main sequence galaxies finds \aco\ within
20\% of the Galactic value][]{BOURNAUD2014a}.

Given the sizes we measure for the molecular emission, the molecular
masses inferred for MD94 and BX610 imply corresponding average
deprojected surface densities of $\Sigma_{\rm
mol}\sim900$~\msunperpcsq\ ($i\sim40^\circ$) and $400$~\msunperpcsq\
($i\sim65^\circ$) assuming the disk size is that of the deconvolved
major axis and the measured elongation is due to projection effects
($\Smol\sim1200-1100$~\msunperpcsq\ in projection). The surface
density found for BX610 is approximately the typical
$\Smol\sim400$~\msunperpcsq\ inferred by
\citet{FREUNDLICH2013} using position-velocity information in four
$z\sim1.2$ galaxies, as well as spatially-resolved measurements in
$z\sim1.5$ galaxies \citep{TACCONI2010,GENZEL2013}. MD94, on the other
hand, exhibits higher deprojected disk surface densities than BX610
mostly as a consequence of its smaller inferred inclination, that
place it at the upper end of the \citeauthor{FREUNDLICH2013}
measurements. We note that these ``deprojected'' estimates are very
tentative and, although more physically interesting than projected
quantities, also considerably more uncertain because of the poor
knowledge of the geometry. In particular, using the H$\alpha$-derived
inclination of BX610 (see next paragraph) would result in a
deprojected $\Sigma_{\rm mol}\sim900$ ~\msunperpcsq.

\begin{figure*}
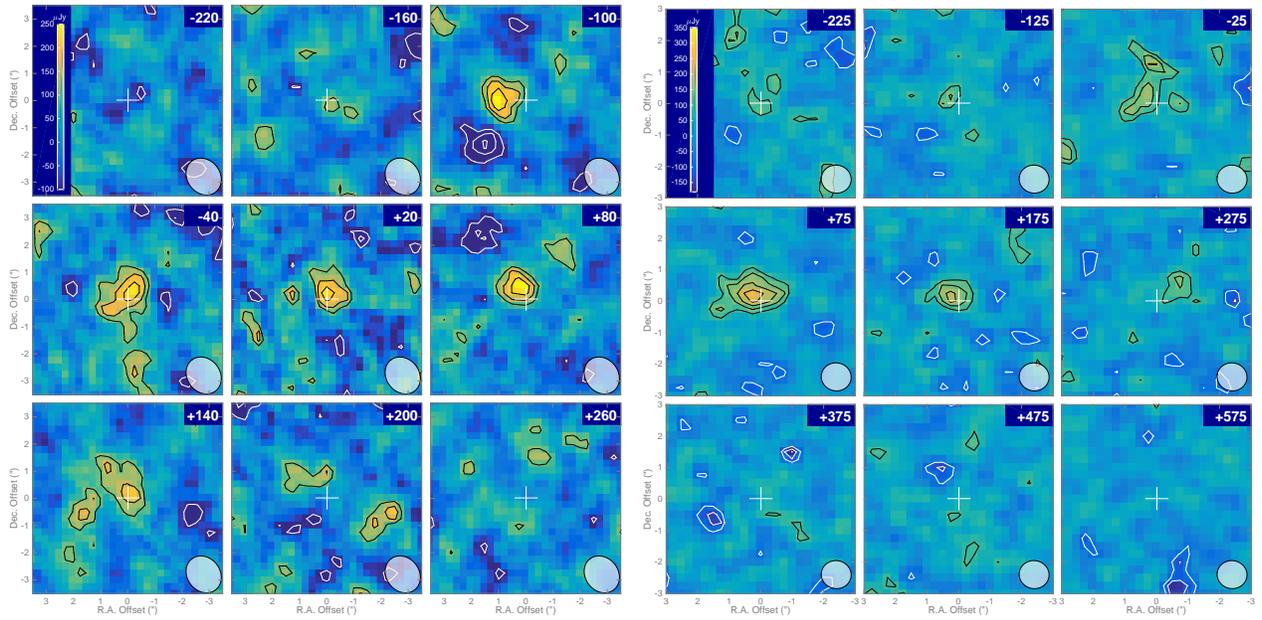

\begin{center}
\includegraphics[width=3.2in]{chmaps_bx610_60.eps}\ \ \ \includegraphics[width=3.2in]{chmaps_md94_100.eps}
\caption{
Channel maps showing the \jone\ emission in BX610 (left) and MD94 (right). We show here the robust combination of C+D data. 
The contours start at $2\sigma$ and have increments of $1\sigma$, with positive values in black and negative in white. 
Note that the velocity interval is $60$~\kmpers\ for BX610 and $100$~\kmpers\ for MD94. The sythesized beams are shown in the
bottom right corner.\label{fig:chmaps}
}
\end{center}
\end{figure*}

We can also extract size and velocity information from the data cubes
themselves, and obtain dynamical mass
estimates. We use GalPak3D, a Bayesian multiparameter Markov Chain
Monte Carlo fitter for 3D galaxy data that takes into account the
effects of instrumental resolution
\citep{BOUCHE2015}, to fit the robust C+D cubes for MD94 and BX610.
The process requires a starting set of assumptions (source center,
size, inclination, velocity dispersion, etc), that are then used to
produce a 3D model that is compared to the data to compute a reduced
$\chi^2$ figure-of-merit which is then minimized. All resulting
estimates are extremely tentative given the resolution and
signal-to-noise of the data (Figure \ref{fig:chmaps}). The reduced
$\chi^2$ resulting from the minimization are 1.15 and 1.05 for MD94
and BX610 respectively, showing that the model is a good fit to the
data. The minimum, however, is shallow, showing that the models are
not unique.  BX610 is fit by a (thick) rotating disk with a maximum
velocity of $v_{rot}\sim220$~\kmpers\ and intrinsic dispersion
$v_{disp}\sim40$~\kmpers, inclination of $i\sim50^\circ$, and radius
$\Rco\sim0.4\arcsec$. These parameters are comparable to those
obtained from the analysis of its H$\alpha$ kinematics
\citep[][Table 2]{CRESCI2009}, although both the rotational velocity
and radius are lower (the latter is also lower than the radius we
obtain from the integrated intensity map), and the inclination is
higher. The resulting dynamical mass computed as $M_{dyn}=2\,\Rco
v_{rot}^2/G$ using the more robust $\Rco=4.1$~kpc estimate from Table
\ref{tab:parameters} 
%and correcting it by $1/\sqrt{\cos i}$ to account for the effect of inclination on the geometry, 
is $M_{dyn}\sim1.1\times10^{11}$~\msun, lower than found by
\citet{CRESCI2009} and too low compared to the inferred baryonic mass
$M_*+M_{\rm mol}\simeq2.1\times10^{11}$~\msun. This is, however, entirely
attributable to the inclination found in the CO fit. If we instead use
the H$\alpha$- derived inclination ($i\sim33^\circ$) we correct
$v_{rot}$ to $\sim310$~\kmpers, resulting in
$M_{dyn}\sim2.0\times10^{11}$~\msun\ in excellent agreement with the
estimate of the baryonic mass (we do not expect dark matter to make a
significant contribution inside $R_{1/2}$). MD94, on the other hand,
is fit as a dispersion-dominated system ($v_{disp}\sim180$~\kmpers\
and $v_{rot}\sim60$~\kmpers\ with $i\sim45^\circ$). The corresponding
dynamical mass is $M_{dyn}\sim0.9\times10^{11}$~\msun, too low to
accommodate even the stellar mass. Unfortunately there are no
H$\alpha$ kinematics to compare with, and the example of BX610
highlights the uncertainties of the analysis.

An interesting local comparison may be provided by the starbursting
region of the nearby galaxy NGC253, which has 50\% of its CO emission
within an area of $150\times50$~pc in projected size with a luminosity
L$_{\rm CO}\approx3.3\times10^8$~\Lcounits\ \citep[][Table
5]{LEROY2014}. That region of NGC253 has a star formation rate
SFR$\sim2-3$ \msunperyr, while MD94 and BX610 have inferred integrated
SFR$\sim271$ and $212$~\msunperyr\ respectively
\citep{TACCONI2013}. About $\sim100$ NGC253-like complexes filling 
$\sim2\%$ of the area of these main sequence disks would be compatible
with our CO luminosity and surface brightness, while simultaneously
fueling the observed star formation activity. This conceptual
``picture'' of an agglomeration of starburst-like regions also fits
the $r_{31}$ excitation as we discuss in the next section. Note,
however, that there is no deep physical significance to the inferred
beam filling area other than the fact that it reproduces the observed
integrated surface brightness.

%Interestingly, this is
%in line with the trend reported by \citet{IONO2009}, who find that
%more luminous LIRGs and ULIRGs in the local universe have smaller CO
%\jthree\ sizes.
 
\subsection{The Excitation of CO}
\label{sec:excitation}

The Cosmic Microwave Background (CMB) has a temperature of
$T_{CMB}\approx9$~K at $z\approx2.3$, which sets an effective floor
for the physical temperature of the cold molecular gas \citep[see
e.g.,][]{DACUNHA2013}.  In terms of the Rayleigh-Jeans brightness
temperature (radiation temperature) measured by the interferometer,
the CMB presents a background that lowers the contrast. Thus even for
optically thick, thermalized emission the observed Rayleigh-Jeans
brightness temperature is systematically lower than the excitation
temperature, $T_{ex}$, and vanishes when $T_{ex}=T_{CMB}$ in the rest
frame \citep[see Eq. 6 in][which is applicable to the rest
frame]{BOLATTO2013}. The fact that we see relatively bright CO \jone\
and \jthree\ emission already says that $T_{ex}\gg9$~K. 

The effect of the CMB on the ratio of two transitions is small unless
the excitation temperature of the emission is very low. The $S_{\rm
CO}\Delta v$ integrated flux ratio between the \jthree\ and \jone\
transitions is $10.6\pm1.5$ for MD94 and $8.3\pm1$ for
BX610. Converting to ratios in brightness temperature units requires
dividing by the ratio of upper level quantum numbers, $J^2$, resulting
in $r_{31}\approx1.17\pm0.17$ and $r_{31}\approx0.92\pm0.11$ for MD94
and BX610 respectively. These data have much better signal-to-noise in
CO \jone\ than observations used by \citet{ARAVENA2014} to measure
$r_{31}=0.58^{+0.21}_{-0.13}$ in BX610, and the $r_{31}$ we obtain is
at the $1.5\sigma$ upper boundary of the previous measurement.

Under the assumption of coextensive emission, these integrated numbers
suggest that the $T_{ex}$ of the \jone\ and
\jthree\ transitions are very similar, and both transitions arise from
optically thick warm gas (that is, gas with excitation and kinetic
temperature significantly higher than the $E_{32}/k\approx16.6$~K energy
necessary to collisionally jump from $J=2$ to the $J=3$ level, where
$k$ is Boltzmann's constant).

\begin{figure*}
\begin{center}
\includegraphics[width=3.5in]{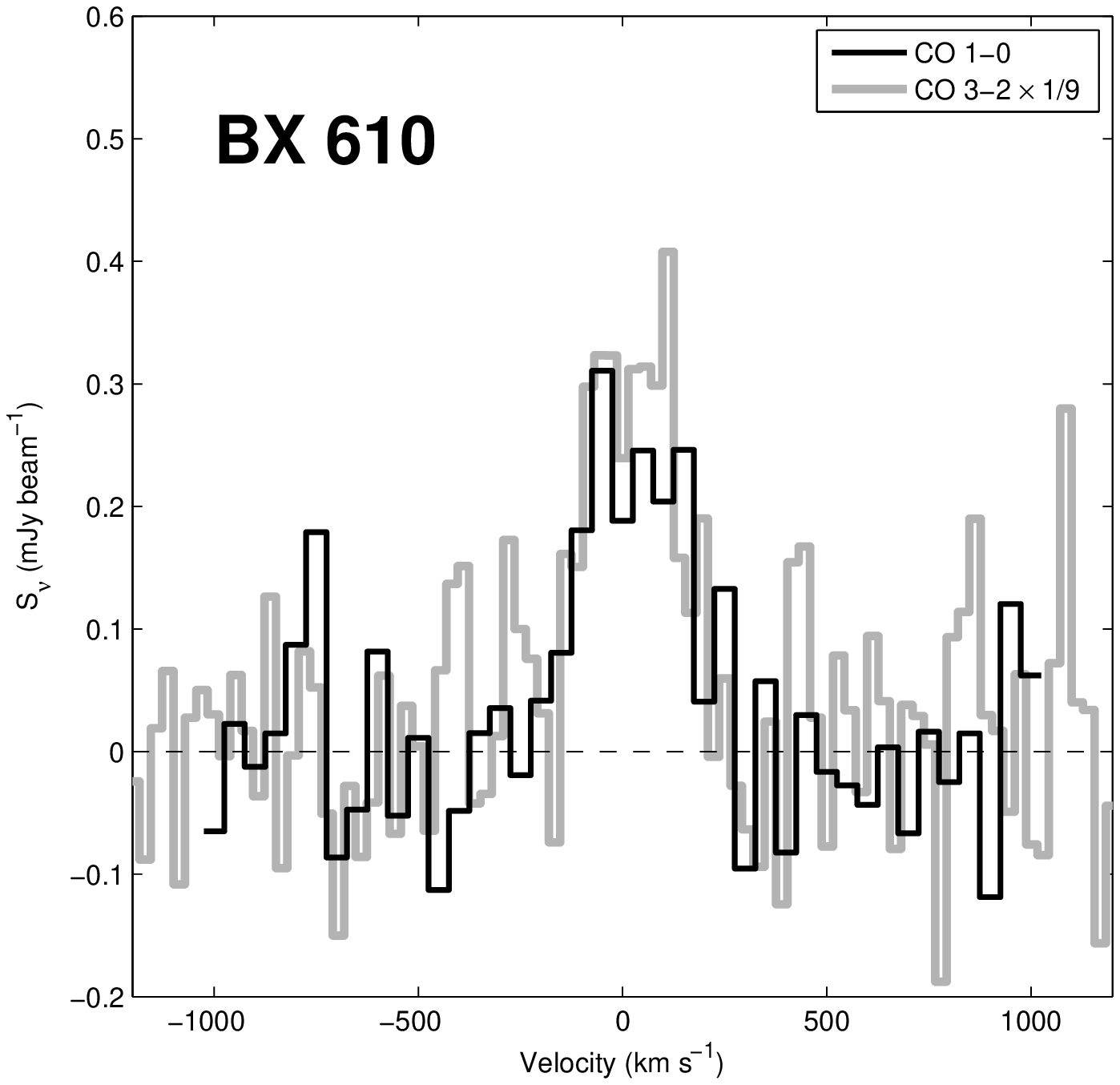}\includegraphics[width=3.5in]{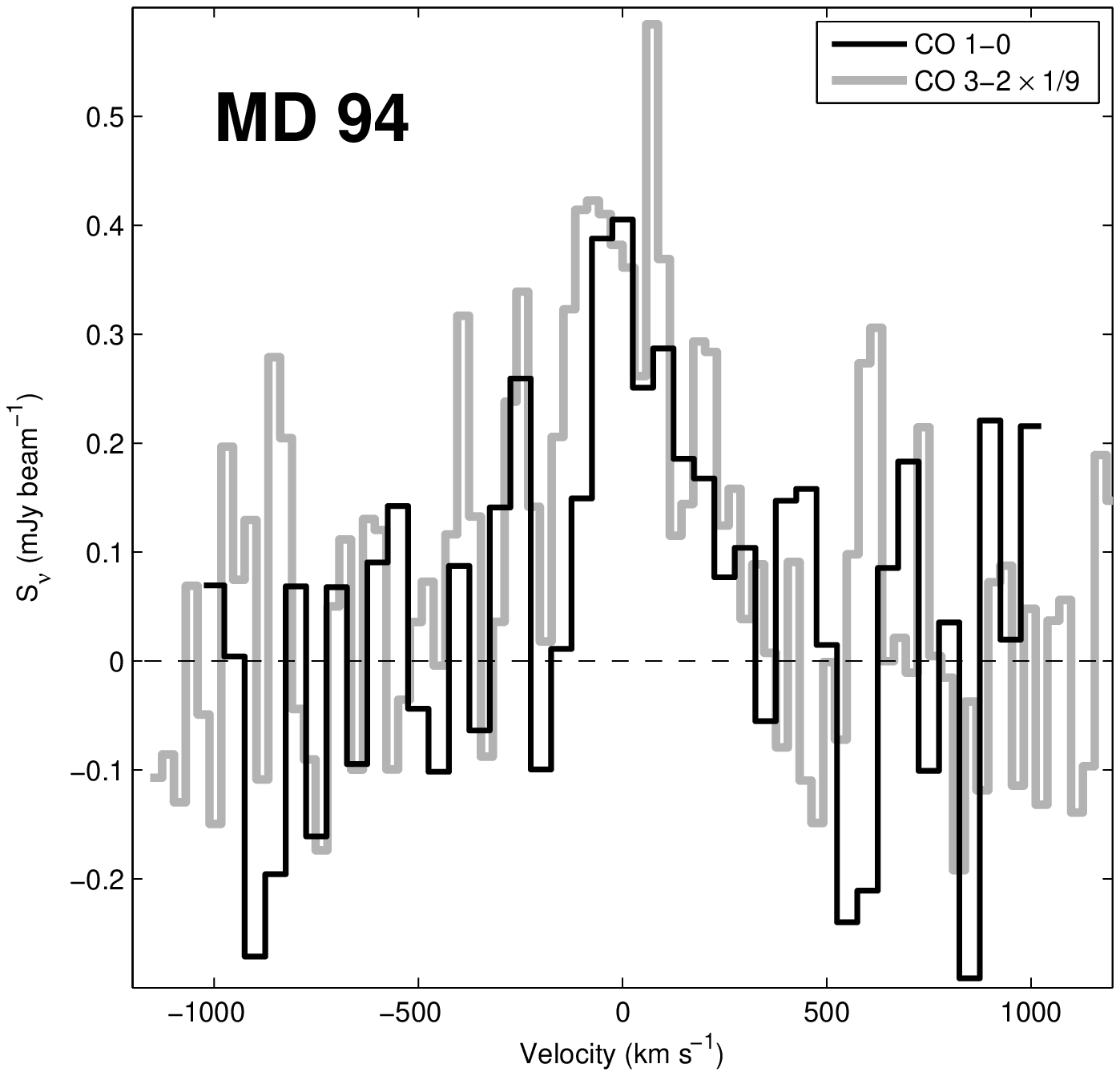}
\caption{
Comparison of spatially integrated spectra of MD94 and BX610 in CO
\jone\ and \jthree. The spectral cubes are matched at $2\farcs5$
resolution for BX610 and $5\arcsec$ resolution for MD94. The spectral
resolution is 50~\kmpers\ for \jone\ and $\sim28$~\kmpers\ for
\jthree. The panels show the spectra for the center position after
spatial resolution-matching. The \jthree\ data (in gray) are scaled down by a
factor of 9 (i.e., $J^2$). \label{fig:spectra}}
\end{center}
\end{figure*}

The data strongly suggest that the emission in CO \jone\ and \jthree\
is coextensive in these objects.  The source sizes are very similar in
both CO transitions in BX610, where the measured sizes of the CO emission
for the \jone\ and \jthree\ transitions are the same within the
uncertainties ($R_{1/2}=4.1\pm1.1$~kpc for \jone,
$R_{1/2}=3.2\pm0.6$~kpc for \jthree). Moreover, the spectral shape of
the emission is very similar for both transitions in MD94 and BX610,
within the uncertainties associated with the noise (Figure
\ref{fig:spectra}). The respective linewidths in CO \jone 
(CO \jthree) from Single Gaussian fitting are $296\pm66$ ($430\pm61$)
\kmpers\ and $294\pm49$ ($266\pm35$) \kmpers\ respectively (the
difference in the case of MD94 is driven by the blue side of the
spectrum, where the fit to \jthree\ incorporates as line a wider range
of velocities resulting in a shift of the centroid). Averaging these
results yields a typical $\Delta v (\jone)/\Delta v
(\jthree)=0.90\pm0.15$ on average for these two sources.  This
suggests that, within our ability to quantify them with the present
data, no dramatic excitation gradients are present, as both
transitions sample the same gas kinematics. This is in contrast with
what may be the situation in SMGs, where \citet{IVISON2011} find a
measurably larger typical linewidth for the CO \jone\ in comparison to
\jthree, suggesting the colder gas is more extended. Note, however,
that this is a subtle $\sim15\%$ effect in the SMG sample, and so it
could go undetected in our data.

These results point to a molecular gas excitation that is high and
uniform in MD94 and BX610, up to the $J=3$ level. Given the widespread
molecular emission, the large average gas surface densities, and the
inferred SFRs, this should not be surprising. The excitation
requirements for the \jthree\ transition are not particularly
stringent: $T\gtrsim E_{32}/k \approx16.6$~K, and an effective critical
density $n_{crit}\approx2\times10^4/\tau_{\rm CO\,\,3-2}\sim {\rm
few}\times10^3$ \percmcu, making the $r_{31}$ ratio a rather blunt
indicator of excitation. By comparison, the average relation for dust
temperature along the main sequence predicts $T_{dust}\sim 31$~K for
these objects \citep[][Table 5]{GENZEL2015}, which can be taken as a
proxy for the gas temperature in dense environments and certainly
meets the requirement for excitation to $J=3$. The measured $r_{31}$ values
show that there is not a large reservoir of cold and/or diffuse low
excitation molecular gas, bright in CO \jone\ but not apparent in
\jthree, present in these sources.

In a study using a library of galaxy simulations,
\citet{NARAYANAN2014} make the case for the link between \Ssfr\ and CO
line ratios in galaxies. They show that in moderately active star-forming
galaxies ($\Ssfr\gtrsim0.2$~\msunperyrkpcsq) the molecular gas has
average temperatures larger than the energy of the $J=3$ level, and
given its large typical optical depth it is very easy to excite this
transition through collisions and radiative trapping. In particular,
they parametrize the expected line ratios as a function of
\Ssfr. Using their Eq. 19 and the $\Ssfr\gtrsim2-3$~\msunperyrkpcsq\ measured
for our galaxies, we would expect $r_{31}\sim0.9$, very similar to
what we measure. 

Interestingly, our results are also consistent with the $r_{31}$
measured for CO in the central regions of NGC253
\citep{BRADFORD2003}. Another point of comparison is provided by the
nearby Seyfert NGC1068.  There the line ratio averages to
$r_{31}\sim1.2$ in the starburst ring, but with a large range of
values correlated with the local star formation activity measured in
Paschen $\alpha$: $r_{31}\sim0.7-1$ in regions with low SFR, and
$r_{31}\sim2-3$ in regions with high SFR
\citep{GARCIA-BURILLO2014}. The compact circumnuclear disk region,
heavily influenced by the AGN, shows $r_{31}\sim2.7$.

\begin{figure*}
\begin{center}
\includegraphics[width=4in]{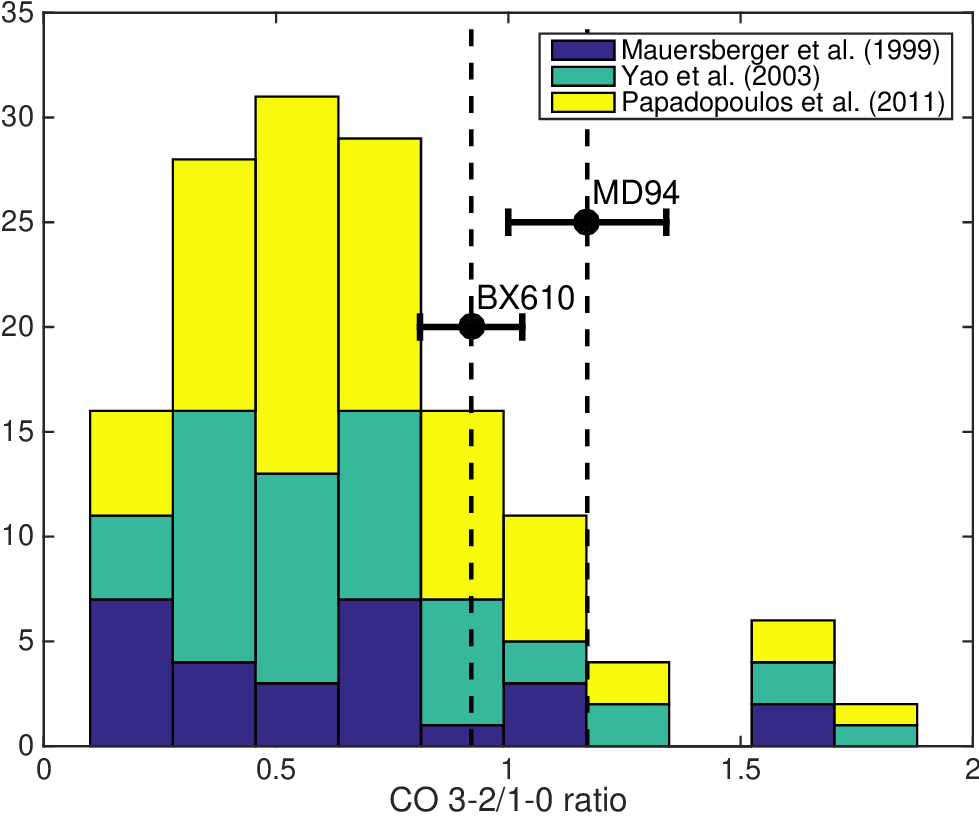}
\vspace{0.2in}
\caption{
Compilation of $r_{31}$ values in local samples, measured in nearby galaxies \citep{MAUERSBERGER1999}, luminous infrared galaxies \citep{YAO2003}, and ultra-luminous infrared galaxies \citep{PAPADOPOULOS2011}. All samples have very similar mean values of $r_{31}\sim0.66$. The values measured here for MD94 and BX610 (c.f., Table \ref{tab:parameters}) fall on the high end of the observed range, but are not outliers. \label{fig:ratplot}}
\end{center}
\end{figure*}

Actively star-forming $z=0$ galaxies span a range of $r_{31}$, and
although a typical value is $r_{31}\sim0.66$ there exist several local
examples with $r_{31}\ge0.9$
\citep[Figure \ref{fig:ratplot};][]{MAUERSBERGER1999,YAO2003,PAPADOPOULOS2011}. There has been a
tendency to attribute the high $r_{31}$ measured in some high-$z$
sources to the presence of a strong AGN \citep[e.g.,][]{RIECHERS2011}.
Nonetheless, it is clear now that at least some high-$z$ sources
display high $r_{31}$ ratios with no clear AGN activity
\citep{SHARON2013,SHARON2015}. That is certainly the case also in
the NGC253 starburst, where no strong AGN is detected, and in the
starburst ring of NGC1068 where the influence of the AGN is
negligible.  It is more likely that $r_{31}$ is not so much acting as
an indicator of overall density and temperature, but rather showing
how well mixed the star formation activity (which provides the bulk of
the gas heating) and the molecular reservoir are. Galaxies where part of
the molecular reservoir is quiescent (thus cold and/or diffuse) will
show lower $r_{31}$, while galaxies where most of the molecular gas in
experiencing star formation will exhibit higher $r_{31}$ ratios.

\section{Summary and Conclusions}

We present and discuss CO \jone\ observations of four main sequence
star forming galaxies at $z=2.2-2.3$ that are part of the PHIBSS
sample. These observations image and resolve two of these galaxies,
MD94 and BX610, with very good signal-to-noise. These galaxies are
representative of the luminous and massive end of the main sequence at
their redshift. The reader should keep in mind that because they were
selected on the basis of their CO \jthree\ flux, however, it is
unclear whether they are representative of the CO properties of the
wider main-sequence population. We use these data to study three
aspects of these galaxies:

\begin{enumerate}
\item We clearly resolve their molecular disks in CO \jone, finding
half-light CO radii of $\Rco\approx4.8$ and 4.1~kpc
respectively. We find that the molecular and optical disk of BX610
have half-light sizes that agree within the errors, while MD94 lacks
high resolution rest-frame optical observations to compare with.
This agrees with the conclusions of
\citet{TACCONI2013}, derived using CO \jthree\ measurements. It is
also similar to the observations on local disks where both CO and
optical light have very similar length scales
\citep{YOUNG1991,SCHRUBA2011}, suggesting that in these $z\sim2$ main
sequence objects we are already observing an equilibrium configuration
between the established stellar population and the CO-bright molecular
gas fueling star formation.  We also observe that the CO \jone\
molecular disk sizes we measure in these main-sequence galaxies are
more extended than those typically obtained for mid-J CO emission in
SMGs at similar redshifts.

\item We measure molecular masses very similar to the stellar masses of 
these galaxies, and gas surface densities that are
$500-900$~\msunperpcsq\ deprojected, assuming that the CO-to-\htwo\
conversion factor is similar to the Milky Way \citep[which is
supported in general for main sequence galaxies,
e.g.,][]{GENZEL2015}. The CO surface brightness is similar to what
would be produced by $\sim100$ complexes like that in the center of
NGC253 filling $\sim2\%$ of the area of these disks. This ``picture''
of a collection of NGC253-like starburst complexes is also consistent
with the observed SFR and $r_{31}$ excitation.

\item Comparison of our observations with \jthree\ data yields $r_{31}$
ratios that are approximately unity. This, together with the fact that
we measure very similar line profiles and disk sizes in both
transitions suggests that these sources do not have a large excitation
gradient or a significant reservoir of cold or low density molecular
gas. Because it is relatively easy to populate the $J=3$ level of CO
in star-forming gas \citep[where star formation supplies the heating
needed to raise the gas temperature, e.g.,][]{NARAYANAN2014}, we
suggest that $r_{31}$ may act as an indicator of how well-mixed the
star formation activity is with the entire molecular reservoir.

\end{enumerate}

\acknowledgements We thank Andy Harris
for enlightening discussions and comments on earlier versions of this
manuscript. We also thank Alice Shapley for providing the HST image of
MD94, which includes perfected astrometry. A.D.B.  wishes to
acknowledge partial support from grants NSF-AST0955836 (CAREER) and
NSF-AST1412419, as well as visiting support from the Humboldt
Foundation and the Max Planck Institutes for Extraterrestrial Physics
and Astronomy.

\begin{deluxetable*}{lcccccccc} 
\tablecolumns{9} 
\tablewidth{0pc} 
\tablecaption{Photometry \label{tab:photometry}
}
\tablehead{ 
\colhead{Dataset} & \multicolumn{2}{c}{Position (J2000)\tablenotemark{d}}    & \colhead{$\sigma$\tablenotemark{e}} &  
\colhead{$S_{\rm CO}\Delta v$ \tablenotemark{d}}    & \colhead{L$_{\rm CO}^\prime$} & \multicolumn{3}{c}{Deconvolved source size\tablenotemark{d,f}} \\
& & & & & & \colhead{Major} & \colhead{Minor} & \colhead{P.A.}\\ 
\colhead{} & \colhead{RA} & \colhead{Dec} & \colhead{(mJy km s$^{-1}$)} & 
\colhead{(mJy km s$^{-1}$)} & \colhead{(10$^{10}$\Kkmperspcsq)} & \colhead{($\arcsec$)} & \colhead{($\arcsec$)} & \colhead{($\arcdeg$)}} \\ 
\hline
\hline
\startdata 
%BX610 C & ${\rm 23^h 46^m 09\fs45}$ & $12\arcdeg49\arcmin19\farcs3$ & 13.2\tablenotemark{a} &  
% $170\pm37$\tablenotemark{a} & $4.2\pm0.9$ & $1.7\pm0.8$& $0.5\pm0.3$ & $30\pm11$ \\
BX610 C & ${\rm 23^h 46^m 09\fs43}$ & $12\arcdeg49\arcmin19\farcs3$ & 14\tablenotemark{a} &  
 $124\pm35$\tablenotemark{a} & $3.1\pm0.9$ & $1.0\pm0.4$& $0.4\pm0.2$ & $27\pm67$ \\
BX610 C+D     & ${\rm 23^h 46^m 09\fs43}$ & $12\arcdeg49\arcmin19\farcs1$ & 7\tablenotemark{a}  &  
$105\pm10$\tablenotemark{a} & $2.6\pm0.2$ & $1.6\pm0.4$& $0.6\pm0.4$ & $118\pm22$\\
BX610 CO 3-2  & ${\rm 23^h 46^m 09\fs44}$ & $12\arcdeg49\arcmin19\farcs3$ & 40\tablenotemark{a}  &  
 $870\pm63$\tablenotemark{a} & $2.4\pm0.2$ & $0.8\pm0.2$& $0.7\pm0.2$ & $34\pm123$\\
MD94 C        & ${\rm 17^h 00^m 42\fs00}$ & $64\arcdeg11\arcmin24\farcs3$ & 14\tablenotemark{b} & 
 $288\pm68$\tablenotemark{b} & $7.8\pm1.5$& $1.6\pm0.4$& $1.0\pm0.3$ & $6\pm20$\\ 
MD94 C+D      & ${\rm 17^h 00^m 42\fs01}$ & $64\arcdeg11\arcmin24\farcs2$ & 10\tablenotemark{b} & 
 $144\pm13$\tablenotemark{b} & $3.9\pm0.4$ & $1.1\pm0.3$& $1.0\pm0.3$ & $114\pm86$\\ 
MD94 CO 3-2   & ${\rm 17^h 00^m 42\fs13}$ & $64\arcdeg11\arcmin24\farcs6$ & 170\tablenotemark{b} & 
 $1520\pm170$\tablenotemark{b} & $4.6\pm0.5$ & \nodata & \nodata & \nodata \\
MD69 C        & ${\rm 17^h 00^m 47\fs67}$ & $64\arcdeg09\arcmin44\farcs3$ & 10\tablenotemark{a} &
 $52\pm18$\tablenotemark{a}  & $1.4\pm0.5$ & $1.3\pm0.6$ & $0.2\pm0.3$ & $81\pm28$ \\
MD174 C        & \nodata & \nodata & 40\tablenotemark{c} &
 $<120$\tablenotemark{c}  & $<3.3$\tablenotemark{c} & \nodata & \nodata & \nodata \\
\enddata
\tablenotetext{a}{Integrated over $\pm150$ \kmpers.}
\tablenotetext{b}{Integrated between $-250$ \kmpers\ and $+500$ \kmpers\ with respect to $z=2.336$.}
\tablenotetext{c}{Integrated over $\pm300$ \kmpers, the width of the \jthree\ emission. Limits are $3\sigma$.}
\tablenotetext{d}{Deconvolved source FWHM diameters from 2D Gaussian fitting using {\tt imfit}.}
\tablenotetext{e}{In a beam, in the integrated intensity map.}
\tablenotetext{f}{For C+D, reported for robust Briggs weighting}.
\end{deluxetable*}

\begin{deluxetable*}{lccccccccc} 
\tablecolumns{10} 
\tablewidth{0pc} 
\tablecaption{Galaxy Parameters \label{tab:parameters}
}
\tablehead{ 
\colhead{Galaxy} & \colhead{Stellar Mass\tablenotemark{a}} & \colhead{SFR\tablenotemark{a}} & \colhead{Molecular Mass\tablenotemark{b}}        & \multicolumn{2}{c}{Diameters\tablenotemark{c}}   & \colhead{\Rco} & \colhead{$I_{\rm CO}$\tablenotemark{d}} & \colhead{$\Sigma_{\rm mol}$\tablenotemark{b,e}} & \colhead{$r_{31}$}\\ 
                 & \colhead{(10$^{11}$ M$_\odot$)} & \colhead{(\msunperyr)} & \colhead{(10$^{11}$ M$_\odot$)} & \multicolumn{2}{c}{(kpc)}  & \colhead{(kpc)} & \colhead{(\Kkmpers)}   & \colhead{(\msunperpcsq)}     &                  }
\hline
\hline
\startdata 
BX610 & $1.0\pm0.3$ & $212\pm74$ & $1.1\pm0.1$ & $12.2\pm3.4$ & $4.6\pm2.5$ & $4.1\pm1.1$ & $248\pm135$ & $398\pm146$ & $0.92\pm0.11$ \\ 
MD94\tablenotemark{f}  & $1.5\pm0.5$ & $271\pm95$ & $1.7\pm0.2$ & $10.8\pm1.2$ & $8.3\pm1.2$ & $4.8\pm0.4$ & $277\pm55$  & $928\pm164$ & $1.17\pm0.17$ \\
\enddata
\tablenotetext{a}{From \citet{TACCONI2013}.}
\tablenotetext{b}{Assuming a CO-to-\htwo\ conversion $\alpha_{\rm CO}=4.36$~\acounits\ which includes the 1.36 Helium correction factor.}
\tablenotetext{c}{Major and minor diameter enclosing 50\% of CO emission computed using {\tt imfit}.}
\tablenotetext{d}{L$_{\rm CO}^\prime/(2\pi\,R_{1/2}^2)$ resolved integrated intensity, projected on the sky.}
\tablenotetext{e}{Very uncertain deprojected values using size from CO major axis. That has a small effect on MD94, and a much larger effect on BX610. Measured ``on the sky'' values are $\Smol\sim1100-1200$~\msunperpcsq.} 
\tablenotetext{f}{Contains an AGN \citep{ERB2006}.}
\end{deluxetable*}

%\begin{thebibliography}{}
\bibliographystyle{apj-hacked}
\bibliography{references}

%\end{thebibliography} 

\end{document}